# Physics Approaches to the Spatial Distribution of Immune Cells in Tumors

Clare C. Yu[1+§], Juliana C. Wortman[1#], Ting-Fang He[2#], Shawn Solomon[2], Robert Z. Zhang[2], Anthony Rosario[2], Roger Wang[2], Travis Y. Tu[2], Daniel Schmolze[3], Yuan Yuan[4], Susan E. Yost[4], Xuefei Li[5], Herbert Levine[5,6], Gurinder Atwal[7], and Peter P. Lee[2+§]

[1]Department of Physics and Astronomy, University of California, Irvine, Irvine, CA 92697;
[2]Department of Immuno-Oncology, City of Hope Comprehensive Cancer Center and Beckman Research Institute, 1500 East Duarte Road, Duarte, CA 91010
[3]Department of Pathology, City of Hope Comprehensive Cancer Center, 1500 East Duarte Road, Duarte, CA 91010
[4]Department of Medical Oncology and Therapeutics Research, City of Hope Comprehensive Cancer Center, 1500 East Duarte Road, Duarte, CA 91010
[5]Department of Bioengineering and the Center for Theoretical Biological Physics, Rice University, Houston, TX 77030
[6]Department of Bioengineering and Department of Physics, Northeastern University, Boston, MA 02115
[7]Cold Spring Harbor Laboratory, Cold Spring Harbor, NY 11724
[#]J.C.W. and T-F. H. contributed equally to this work.
[*]Corresponding author: cyu@uci.edu

The goal of immunotherapy is to enhance the ability of the immune system to kill cancer cells. Immunotherapy is more effective and, in general, the prognosis is better, when more immune cells infiltrate the tumor. We explore the question of whether the spatial distribution rather than just the density of immune cells in the tumor is important in forecasting whether cancer recurs. After reviewing previous work on this issue, we introduce a novel application of maximum entropy to quantify the spatial distribution of discrete point-like objects. We apply our approach to B and T cells in images of tumor tissue taken from triple negative breast cancer (TBNC) patients. We find that there is a distinct difference in the spatial distribution of immune cells between good clinical outcome (no recurrence of cancer within at least 5 years of diagnosis) and poor clinical outcome (recurrence within 3 years of diagnosis). Our results highlight the importance of spatial distribution of immune cells within tumors with regard to clinical outcome, and raise new questions on their role in cancer recurrence.

# Introduction
## Why physicists?

Why should physicists pay attention to cancer as a research topic? Traditionally, cancer research has been the purview of biologists and medical researchers. Yet, despite the billions of dollars that have been spent on the war on cancer, far too many people are still battling this disease. Consider the following statistics. Worldwide, in 2012, approximately 14 million new cases of cancer were diagnosed and 8 million people died of cancer (15% of deaths) [1]. In the United States alone about 600,000 people die of cancer each year [2]. This accounts for 1 in 4 deaths. It is estimated that during their lifetime, one in two men and one in three women will be diagnosed with cancer [3]. These statistics highlight the need for new ways of thinking. As a result, interdisciplinary collaborations involving cancer researchers, physicists, mathematicians, engineers and computer scientists have been working to hasten progress with new techniques and different approaches.

To compare and contrast the types of problems that physicists and biologists study, consider a typical biological approach that emphasizes signaling pathways. (A signaling pathway is a sequence of switches in a cell in which protein A activates or deactivates protein B that in turn switches protein C on or off, etc. Proteins are molecular machines that perform various jobs in the cell.) Cancer research typically focuses on signaling pathways that have been over-activated or deactivated due to gene mutations. These aberrant signaling pathways enable a tumor to grow and spread. The goal is to develop therapeutics targeting these pathways on which the growing tumor depends. Yet, all too often these chemotherapy drugs, combined with surgery and radiation, provide only a temporary fix as the cancer develops resistance and recurs.

In recent years, a paradigm shift has occurred. Rather than using drugs to kill the cancer, immunotherapy gets the immune system to kill cancer cells. Immunotherapy tends to work better when the immune cells recognize the cancer and infiltrate the tumor. In fact, for some types of cancer, the prognosis is better when the density of killer T cells (that can kill cancer cells) is higher in the tumor, even if no immunotherapy is employed. We will discuss this more below, but suffice it to say that while immunotherapy has shown promising results, we are still a long way from curing cancer.

As Robert Austin has pointed out [4], such failure means that we just do not understand the basic principles behind cancer. We must think more broadly and physicists can help to expand the way we approach cancer. For example, an exclusive focus on signaling pathways and gene mutations ignores the physical characteristics of tumors. In particular, it ignores the spatial aspects of tumors, e.g., the spatial organization of cells and structures. Such spatial structure can affect the ways cells interact with each other and the structures in their environment. Studies on the spatial aspects of tumors could lead to different approaches and treatments.

In this article, after an overview of cancer, immunology and immunotherapy, we give examples of how techniques used by physicists can be applied to quantify the spatial distribution of immune cells in tumors. These techniques include fractal dimensions and maximum entropy.

**Epidemiology**

Worldwide, the most commonly diagnosed cancers are lung, breast and colorectal cancer, and the most common causes of cancer death are lung, liver and colorectal cancer [1]. In the United States the most common cancers occur in the breast, lung and prostate, and the most common causes of cancer death are lung, colorectal and pancreatic cancer [5].

**What are the causes of cancer?** Cancer risk factors include genes and family history. (A gene is a sequence of DNA bases that code for a corresponding sequence of amino acids that fold to form a protein molecule [6].) However, some cancers, such as breast and prostate cancer, are genetically heterogeneous. This means that different gene mutations are found in tumors from different patients with nominally the same type of cancer [7, 8]. For example, a study of 510 breast tumors from 507 patients found 30,626 gene mutations, but mutations in only three genes occurred in more than 10% of the tumors [8].

There are environmental factors such as carcinogens (cancer causing substances), radiation, and sunlight [5] that increase cancer risk. Lifestyle matters. Tobacco, alcohol, a diet lacking fruits and vegetables along with excess salt all increase the risk of cancer [5, 9]. Obesity is a risk factor along with diabetes [5, 10-12]. In addition, certain occupations have an increased cancer risk, e.g., those with exposure to carcinogenic chemicals or asbestos [5]. People working the night shift may also have an increased risk of cancer [13-16].

Infectious agents, i.e., viruses and bacteria, were associated with 19% of cancers in 2002 [17]. Examples include strains of the human papilloma virus (HPV) that can cause cervical cancer [18, 19], hepatitis B and C which increase the risk of liver cancer possibly through chronic inflammation [20, 21], HTLV-1 which can cause leukemia [22], *Helicobacter pylori* which is the primary cause of stomach ulcers and increases the risk of stomach cancer through chronic inflammation [23], and the Epstein-Barr virus which can cause Burkitt's lymphoma and nasopharyngeal (nose and throat) cancer [24]. Chronic inflammation increases the risk of cancer because inflammation involves inflammatory cells and growth factors (substances that promote cell growth) that can encourage cancer cells to proliferate [25, 26].

**Cancer mortality**
In spite of the large amount of time, money and effort that has gone into battling cancer, cancer death rates have not declined much in the United States. From 2004 to 2013, the overall cancer death rate in the US fell by 13% [2]. However, between 2012 and 2030, worldwide cancer death rates are projected to increase by 60% from 8 million to 13 million per year [5].

**What can be done to lower cancer mortality?**
Prevention is always the best choice. Having a healthy lifestyle lowers the risk. Avoid smoking, eat healthy, and exercise regularly. Cancer vaccines would be wonderful given how effective vaccines have been in preventing diseases caused by microorganisms. So far, though, the only cancer vaccine in widespread use is Gardasil which vaccinates against the strains of HPV that commonly cause cervical cancer [27] and some types of oropharyngeal (throat) cancer [28, 29].

Early detection is key for successfully surviving cancer. Examples of cancer screening include the PSA (prostate serum antigen) blood test for prostate cancer, mammograms for breast cancer, and colonoscopies for colorectal cancer. Colonoscopies can actually help to prevent cancer of the colon and rectum by removing polyps that can later become cancerous.

New technologies and approaches could lead to novel treatments. Recently there have been efforts in both the public and private sector to encourage physicists, mathematicians and engineers to collaborate with cancer researchers and clinical oncologists (medical doctors specializing in cancer). For example, the National Cancer Institute sponsors a Physical Sciences Oncology Network, and Stand Up to Cancer sponsors interdisciplinary convergence teams consisting of theoreticians and oncologists. Since cancer mortality is still high, the hope is that new perspectives and approaches from other fields could lead to new discoveries.

**Limitations on Current Cancer Research**
Scientific advances through research can lead to significant progress in the fight against cancer, but progress is slow. While new drugs have resulted from cancer research, there is the caveat that about 90% of preclinical cancer research results are not reproducible [30]. The biotech company Amgen tried to confirm the results of fifty-three landmark studies and found that they could only confirm the scientific findings in six cases (11%). In the cases where the results could not be reproduced, the investigators often presented the results of only one experiment rather than findings that were reflective of the entire data set. Amgen's findings are consistent with other studies. For example, Bayer HealthCare in Germany reported that only about 25% of published preclinical studies could be validated [31]. 70% of the studies analyzed by Bayer involved cancer research, some of which might have also been analyzed in the Amgen study.

More exchange of research information before publication could also help. For example, in physics, there is an online preprint archive (https://arxiv.org) where researchers post preprints before publication. Unfortunately, even though such an archive exists in biology (www.biorxiv.org), many biologists are not interested in utilizing such an archive [32]. One reason is that some biologists are afraid of other researchers stealing their results, which is rather ironic given the poor track record of reproducibility of biological research results [30, 31].

**Introduction to Breast Cancer**
Since the rest of this article will focus on breast cancer, we begin with a brief introduction to the biology, structure and function of the breast [33, 34], or mammary gland, which is composed of both glandular and fatty tissue on top of the chest muscles. The function of the mammary gland is to produce milk. Milk is produced in small grapelike sacs called acini. A cluster of acini is called a lobule and a cluster of lobules is called a lobe. The milk produced in acini flows into a small channel called a ductule. Ductules merge to form milk ducts. Each lobe has one milk duct. If you look at the cross-section of a milk duct (see Figure 1), it is lined with mammary epithelial cells. (In general, epithelial cells line the inner surfaces of tubes and cavities in the body, e.g., the lungs and gastrointestinal tract. Carcinomas are cancers that originate from epithelial cells. The space inside the tube or cavity is called the lumen.) Surrounding the mammary epithelial cells is a layer of myoepithelial cells that help to squeeze the milk down the duct. Outside the myoepithelial cells is a tough outer sheath called the basement membrane. It is an example of extracellular matrix and is composed of proteins such as collagen. (Collagen is the most common protein in your body. It helps give elasticity to your skin.)

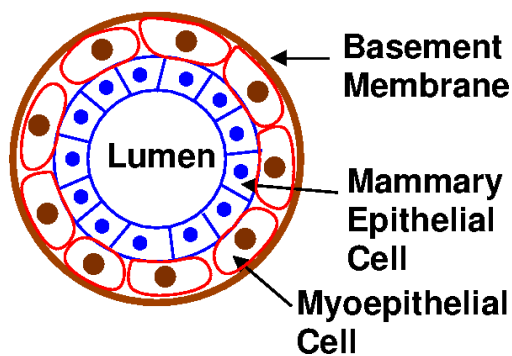

**Figure 1 (color online): Cross section of a normal milk duct showing the lumen, mammary epithelial cells, myoepitheial cells, and basement membrane.**

Most breast tumors start in the milk ducts or lobules. Typically the mammary epithelial cells start to proliferate and fill up the luminal space in the duct or acini. When this occurs to a substantial degree, this is known as ductal carcinoma *in situ* (DCIS). This is stage zero cancer. The phrase "in situ" literally means "in place". In this case, it means that the cells have not broken through the basement membrane. If the epithelial cells continue to proliferate and break out of the duct and through the basement membrane, then this is called invasive ductal carcinoma. "Invasive" means that the cells have invaded tissue where they do not belong. Since invasive cancer cells break through a membrane, tumors are "wounds that do not heal" [35, 36].

Each year in the United States, over 200,000 women are diagnosed with invasive (stage 1 and higher) breast cancer; about 60,000 women are diagnosed with *in situ* (stage 0) breast cancer; about 40,000 women will die of breast cancer; and about 2000 men are diagnosed with breast cancer [5].

One of the more aggressive types of breast cancer is triple negative breast cancer (TNBC). Triple negative breast cancer cells lack appreciable expression of hormone epidermal growth factor receptor 2 (HER-2), estrogen receptors (ER) and progesterone receptors (PR). TNBC occurs in 15-20% of breast cancer cases. According to a recent study of TNBC, 76.2% of recurrences occurred within the first 5 years after surgery, with the median time of recurrence of 2.7 years [37]. We mention this type of breast cancer because some of the results below involve TNBC patients.

## Importance of the Tumor Microenvironment

We tend to think of tumors as consisting solely of cancer cells, but in fact, tumors need a supporting environment known as the tumor microenvironment (TME) in order to survive and thrive [38]. The tumor microenvironment consists of blood vessels, various types of immune cells, fibroblasts, extracellular matrix, etc. Fibroblasts are cells that synthesize extracellular matrix and fibers such as collagen; fibroblasts are the most common cells in connective tissue.

The evolution and progression of cancer depends on the tumor microenvironment. For example, the TME can play an important role in tumorigenesis, i.e., in the formation of tumors [39]. In particular, damage to the tumor microenvironment can cause pre-malignant cells to become malignant [39-42]. For example, injecting a cancer-causing virus (Rous Sarcoma Virus or RSV) into the wing of a chicken produces a tumor at the injection site but not elsewhere, even though the virus has spread throughout the chicken. If the other wing is then wounded, a tumor is formed at the site of the wound [40]. In another experiment [41], pre-malignant cells (COMMA-D mammary epithelial cells with p53 mutations) placed on previously irradiated tissue (mammary fat pads in mice) were about four times more likely to form tumors than when the cells were placed on non-irradiated tissue. Furthermore, the tumors on the irradiated tissue were significantly larger and grew more quickly than those on the non-irradiated tissue. One reason that damaging the tumor microenvironment promotes tumorigenesis is that the growth factors (signaling proteins like TGF-β) that are produced as part of the wound-healing process also promote the inception and growth of tumors [43] as well as metastasis [44].

Lymph vessels are another component of the TME and can be thought of as a sewage system for cells. Cells are bathed in interstitial fluid from which they absorb nutrients and into which they dump waste. Lymph vessels take up interstitial fluid and conduct it to lymph nodes. The fluid inside lymph vessels is referred to as lymph. Eventually lymph is conveyed to blood vessels (veins). The lymphatic system is much more than just waste removal system; it is an important part of the immune system as we describe below.

**Brief Introduction to Lymphocytes**

Immunology is a fascinating and complex field of study. A good introductory text is [45]. Here we will only give a few brief facts to equip the reader with what will be needed to follow this article. White blood cells are immune cells. In this article we will focus on lymphocytes which are a subset of white blood cells. The two main categories of lymphocytes are T cells and B cells. B cells are best known for producing antibodies, which are proteins that recognize and attach to antigens, e.g., a specific protein on invading bacteria or viruses. The antibody tags the invader for destruction by other cells in the immune system, e.g., macrophages. However, B cells can perform other functions such as secreting signaling molecules called cytokines, and presenting antigens for T cells to inspect.

T cells only recognize antigens in the form of pieces of proteins called epitopes that are displayed on the surface of cells in "display cases" that are called major histocompatibility complexes (MHCs). You can think of an MHC as a hot dog bun and the epitope as the hot dog [45]. The surfaces of T cells have T cell receptors (TCRs) which are antibody-like protein complexes that recognize and bind to specific antigens, i.e., specific epitopes in the MHCs. There are three main types of T cells: killer T cells (cytotoxic lymphocytes), helper T cells, and regulatory T cells. Cytotoxic T cells are able to kill unwanted cells, e.g., cancer cells and virus-infected cells. Helper T cells (Th cells) secrete cytokines which are chemical signaling molecules that direct the actions of other immune cells. Regulatory T cells suppress the immune system in order to keep it from overreacting or from acting inappropriately. In the lab, these different types of T cells can be distinguished by specific protein complexes on their surfaces. In particular, cytotoxic T cells and helper T cells express CD8 and CD4 co-receptors, respectively, on their surface. So we sometimes refer to killer T cells as $CD8^+$ T cells, and helper T cells as $CD4^+$ T cells. The plus sign in the superscript means that these cells stain positive for these protein markers by binding with their respective color-coded antibodies in the immunostaining process.

The lymph nodes are small bean-shaped structures where immune cells, such as T cells, B cells and antigen presenting cells (APCs) that present antigens to certain lymphocytes such as T cells, meet and communicate. B cells and T cells that find their cognate antigens in lymph nodes can also be activated there.

## Brief Introduction to Immunotherapy

Currently, there is a great deal of interest in using immunotherapy to treat cancer by activating the immune system to kill cancer cells. We describe below some ways in which this is being done. (For more details, see [45].)

Monoclonal antibodies are antibodies produced in the laboratory that target specific proteins on the surface of cells. For example, some particularly aggressive types of breast cancer cells express high amounts of a growth factor receptor called Her2. When growth factor proteins bind to (or ligate) this surface receptor, the cancer cells proliferate. The monoclonal antibody Herceptin (trastuzumab) can bind to the Her2 receptor and prevent the growth factor proteins from ligating the receptor, thus blocking the growth signals and slowing growth.

## Checkpoint Inhibitors: Antibodies that block the ability of cancer cells to turn off killer T cells

Checkpoint inhibitors are monoclonal antibodies that are used to interrupt the way that cancer cells deactivate killer T cells. Here is how they work. Activated killer T cells express "checkpoint" protein receptors on their surface that prevent killer T cells from being overly active. For example, one of these checkpoint proteins is programmed death 1 (PD-1). When the ligand for PD-1, which is called PD-L1, binds to PD-1, the killer T cell does not function very well. Cancer cells with PD-L1 basically have a 'key' or ligand can effectively neutralize killer T cells by inserting the PD-L1 'key' into the 'lock' or receptor. Thus the cancer cells can prevent the killer T cells from killing the cancer cells. Checkpoint inhibitors are monoclonal antibodies that bind to, e.g., PD-1 or PD-L1, thus preventing the cancer cell from deactivating the killer T cell. It is analogous to taping over the keyhole or wrapping tape around the key so that the key cannot go into the keyhole. Checkpoint inhibitors prevent PD-L1 from binding to the PD-1.

## CAR T Cell Therapy: Engineering T cell receptors

As we mentioned earlier, T cells have T cell receptors that bind to specific antigens. The TCRs on a given T cell are all the same and recognize one specific antigen which is referred to as the TCR's cognate antigen. Different T cells have different TCRs. If the TCRs on a T cell recognize and bind to their cognate antigen, the T cell proliferates, i.e., makes identical daughter cells. CAR T cell therapy is a rapidly emerging cancer treatment in which the TCR is artificially engineered to recognize antigens on cancer cells. CAR stands for "chimeric" antigen receptor. (Chimera was a beast in Greek mythology that had the head of a lion, the body of a goat and a serpent's tail.) The receptors on CAR T cells are designed to recognize and bind to specific surface proteins. It is important to note that these CAR T cognate antigens need not be presented by MHC molecules [45]. Cells that display these cognate proteins are then destroyed by the CAR T cells. CAR T cell therapy has been most successful in treating blood cancers such as leukemia and lymphoma.

## Spatial Distribution of Immune Cells

Even for patients that do not receive immunotherapy, a high density of tumor infiltrating lymphocytes (TILs) is associated with a good prognosis in several types of cancers [46-48]. For example, higher densities of $CD3^+$ and $CD8^+$ T cells were associated with a lower rate of recurrence in colorectal carcinoma [49]. (CD3 is a marker for all T cells and CD8 is a marker for cytotoxic (killer) T cells.) In the case of patients with triple negative and HER2-positive breast cancers, a higher density of TILs is associated with a better prognosis [50]. (Triple negative breast cancer cells test negative for hormone epidermal growth factor receptor 2 (HER-2), estrogen receptors (ER) and progesterone receptors (PR).) While the prognostic value of the density of $CD3^+$ and $CD8^+$ T cells infiltrating tumors is well known, the clinical significance of other types of immune cells, such as B cells, is less clear [51, 52].

Averaging cell densities over the entire tissue overlooks the spatial heterogeneity in the distribution of TILs within the tumor which may be clinically important [53-55]. If we view the tumor microenvironment in ecological terms, interactions between the different components of this ecosystem depend upon their spatial organization. Regional differences in selective pressures produce microhabitats resulting in phenotypic and genetic heterogeneity [56, 57]. So it is worth considering whether the spatial distribution of TILs is associated with a difference in clinical outcome, i.e., in whether or not the cancer recurs.

## Review of Previous Work on the Spatial Distribution of Immune Cells

This leads to the more general problem of quantifying the spatial distribution (or arrangement) of cells in histology-based images with a single scalar number. There have been a number of efforts to quantify spatial heterogeneity of the tumor microenvironment based on comparing populations of cells [58]. For example, the Morisita-Horn index was used to quantify the spatial colocalization of tumor and immune cells, and it was found that significant colocalization was associated with a higher disease-specific survival in Her2-positive breast cancers [58, 59]. The Getis-Ord analysis [60] was used to locate immune hotspots where the clustering of immune cells was significantly above background. A combined immune-cancer hotspot score was found to be associated with good prognosis in ER-negative breast cancer [61]. A quantitative measure of the infiltration of immune cells into a tumor is the intratumor lymphocyte ratio (ITLR) which is defined as the ratio of the number of intratumor lymphocytes to the total number of cancer cells in a histological sample [62]. A high ITLR was found to be associated with good disease specific survival in ER-negative/Her2-negative breast cancer [62].

Natrajan *et al.* [63] quantified the spatial heterogeneity in breast tumors with regard to the proportions of different cell types, e.g., cancer cells, lymphocytes and stromal cells, in different regions of the tumor by calculating the Shannon entropy and using Gaussian mixture models to fit the distribution of Shannon entropies. Their ecosystem diversity index (EDI) was the number of Gaussians needed to fit the distribution. They found that high EDI values were associated with high micro-environmental diversity and poor prognosis. Somewhat ironically, if most of the regions have high Shannon entropies such that a single Gaussian can be used to fit the distribution, then the EDI is low.

Fractal dimensions [64] have been used to characterize the irregular morphology of tumors [65-67] and vasculature [68, 69] as well as subcellular structures such as mitochondria [70] and nuclei [71]. There are numerous ways to calculate fractal dimensions. In the box counting method, the number $N$ of squares (each with area $L^2$) needed to cover the 2D image of a tumor, say, is proportional to $L^{-d}$, where d is the fractal dimension in the limit that L goes to zero (or a very small value). The more irregular the shape, the higher the fractal dimension is and the poorer the prognosis [65, 66].

Assuming that the structure of tumor tissue is reflected in the arrangement of cancer cell nuclei, Waliszewski *et al*. calculated several different fractal dimensions as well as the Shannon entropy and lacunarity to characterize the spatial distribution of cancer cell nuclei in prostate tumor tissue and compared the results to the corresponding Gleason scores in an attempt to find a more objective way to classify prostate tumor tissue [72, 73].

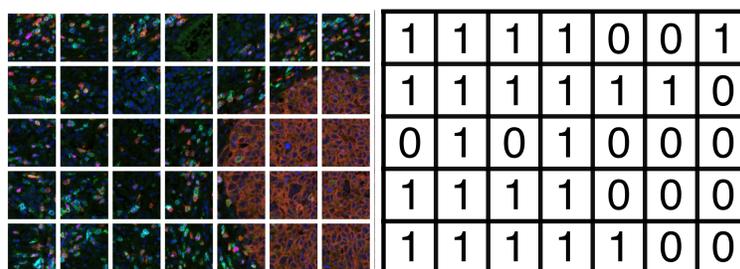

**Figure 2 (color online):** Grid of squares overlaying a tissue image. 1's (0's) correspond to yes (no) answers to the question asked of each square, e.g., "Is there at least one B cell in the square?"

Note that these approaches produce measures of the spatial distribution at a single length scale. More recently, to move beyond this limitation, we previously developed several unique statistical approaches that use coarse graining to examine the spatial distribution of various types of cells and structures within the tumor microenvironment over a range of length scales [74].

<u>Occupancy</u>: The first technique is called 'occupancy'. We begin with an image of tissue where different types of cells are labeled, i.e., immunostained, with different colors, e.g., killer T cells are bright red, B cells are cyan, etc. We overlay the tissue image with a grid of squares as in Figure 2. For each square, we then ask a binary (yes-no) question, e.g., "Is there at least one $CD20^+$ B cell in the square?" If the answer is yes, we assign a '1' to that square. If the answer is no, we assign a '0' to the square. The *occupancy g* is the fraction of squares with 1's, i.e., it is an estimate of the probability that a square will have a 1. To characterize the spatial distribution at different length scales, we varied the size of the squares in the grid and computed the occupancy as a function of L, the length of one side of a square. Note that the occupancy will be affected by the average cell density for questions such as "Is there at least one $CD20^+$ B cell in the square?" because the higher the average cell density, the higher the probability that a square is assigned a '1'. We can plot the occupancy averaged over patients or samples versus the square size L (see Figure 3). We found that the area under the curve (AUC occupancy) differed between triple negative breast cancer patients with good and poor clinical outcome for $CD20^+$ B cells and $CD8^+$ T cells. We defined patients with no recurrence within 5 years after surgery as good clinical outcome (n=24) and patients who had recurrence within 3 years after surgery as poor clinical outcome (n=13).

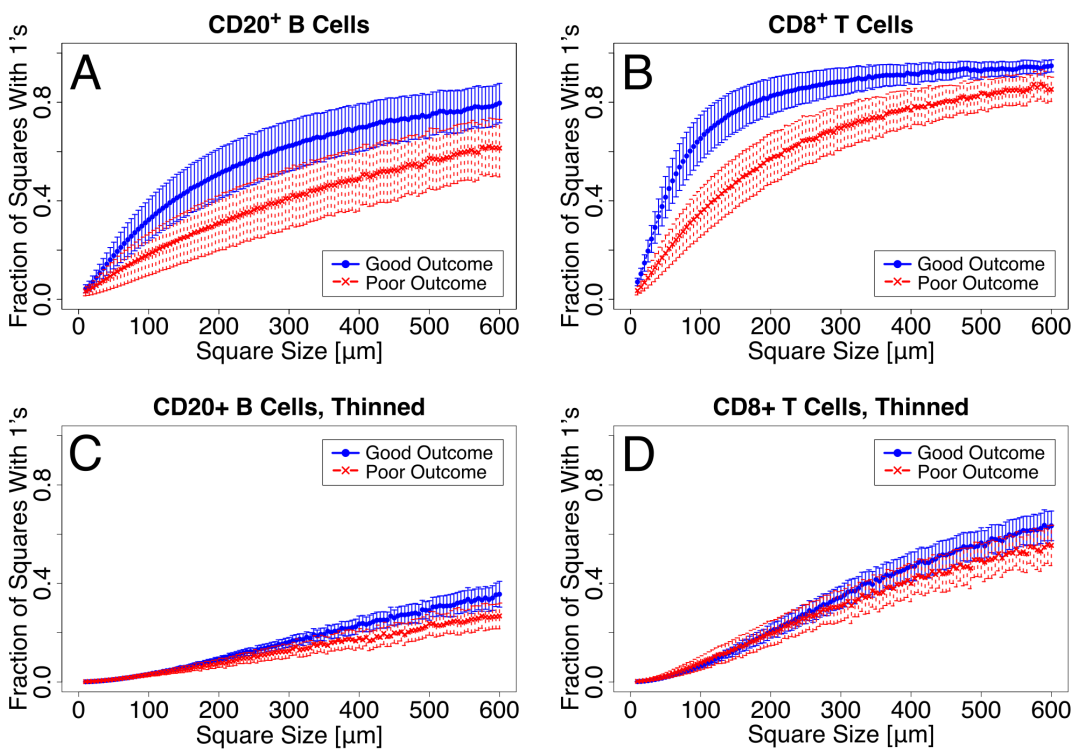

**Figure 3 (color online): Occupancy for $CD20^+$ B cells and $CD8^+$ T cells vs. square size. Notice the difference between good (blue) and poor (red) clinical outcome. The error bars indicate 95% confidence intervals. (A) $CD20^+$ B cell occupancy before thinning (B) $CD8^+$ T cell occupancy before thinning (C) $CD20^+$ B cell occupancy after thinning the B cells to a density of 12 B cells/mm$^2$ (D) $CD8^+$ T cells occupancy after thinning the T cells to a density of 25 $CD8^+$ T cells/mm$^2$.**

Thinning: As we mentioned above, occupancy tends to increase with the average cell density. One way to remove the effect of density on the occupancy is to reduce, or thin, the density by randomly eliminating cells, e.g., $CD8^+$ T cells, in the various images until the densities in all the tissue images have the same value as the image with the lowest density. Here is a simple example of how to randomly remove B cells. To randomly remove half the B cells from an image, you would go to each B cell, flip a coin, and remove the cell if you get 'heads' and keep the cell if you get 'tails'. Figs. 3C and 3D show thinned plots of occupancy vs. square size L. The difference in the area under the curves (AUC) between good and poor clinical outcomes is statistically significant for B cells (p-value=8 x $10^{-4}$) but not for killer T cells (p-value = 0.3). This suggests that the spatial distribution, rather than the density, of B cells differs significantly between good and poor outcomes.

p-value: In the previous paragraph we calculated the p-value under the null hypothesis to ascertain whether a quantity, such as the area under the curve, is clinically significant. The p-value is a standard statistical measure of a binary classifier [75]. Binary refers to our assumption that the clinical outcome is either good or poor. The p-value is the probability of obtaining a random sample with a mean at least as far from that of the null hypothesis as is observed, assuming the null hypothesis is true. In our case, the null hypothesis states that the result could arise by random chance. The smaller the p-value is, the greater the probability that the null hypothesis is not valid. In general, results are considered significant if p<0.05.

Fractal Dimension: Another way to characterize the number of boxes with 1's and hence, the spatial distribution of cells, is with fractal dimension. Although a number of studies have used fractal dimension to analyze morphologies associated with tumors [65-71], we used it to quantify the spatial distribution of individual immune cells. While there are a number of different ways to define the fractal dimension, we chose to use a variation of the box counting method [76]. The number n(L) of squares with '1' will be proportional to $(1/L^\delta)$ where δ is one type of fractal dimension. The constant of proportionality depends on the size of the tissue that dictates the number of boxes covering the tissue. To avoid this, we found it convenient to define the fractal dimension *s* to be given by: *s*(L) = -d[log n(L)]/d[log L]. (Note that unlike the more common definition of the box-counting fractal dimension, we do not take the limit L→0, because we are interested in the distribution of individual cells at different length scales.) We used a different variable, *s*(L), rather than δ, since n(L) typically does not follow a simple power law as L is varied.

Both occupancy and fractal dimension depend on the number of squares with 1's, so there must be a simple relation between the dependence of occupancy on L and fractal dimension d. To find the relation, consider the following. Suppose the total number N(L) of squares covering the image of the tissue goes as $(1/L^D)$. Then if $n(L) \sim (1/L^\delta)$, the occupancy $g = n(L)/N(L) \sim L^{D-\delta}$. Note that D need not be equal to 2 since the image of the tissue may be irregular or there may be regions that were not imaged.

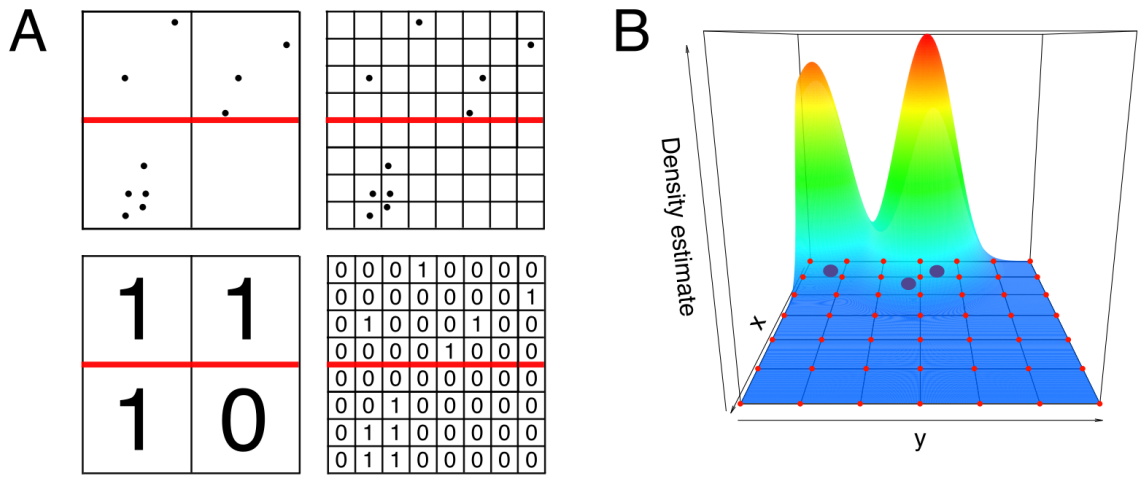

**Figure 4 (color online): (A) Diagram illustrating how fractal dimension (FD) difference can distinguish the difference between spread out cells and clustered cells. The upper halves (above the red lines) of the two images show points that are spread out while the lower halves show clustered points. At long length scales (big boxes) the FD is 2 in the upper half but not in the lower half. At small length scales (small boxes), there is the same number of boxes with points and hence the same fractal dimension. The fractal dimension difference is greater in the upper half because the points are more spread out in the upper half. (B) The Gaussian method for density estimation. Diagram illustrates hotspot analysis. Three large blue points indicate the locations of three cells. Each cell's contribution to the local density is represented by a Gaussian distribution. The mountain over the three cells represents the sum of their Gaussian weights, i.e., the local cell density. The small red points are the grid points where the Gaussian weights are summed.**

Fractal Dimension Difference: To see if cells are clustered or spread out, we looked at the difference $\Delta s$ in fractal dimension between large and small length scales: $\Delta s = s_{Large} - s_{small}$, where $s_{Large}$ is the fractal dimension at large length scales and $s_{small}$ is the fractal dimension at small length scales. The small and large length scales should roughly bracket the typical, or median, nearest neighbor distance between cells of the same type, e.g., $CD8^+$ T cells. In all the cases we examined, $\Delta s > 0$. If $\Delta s$ is large, it means that the cells are more dispersed, i.e., more spatially spread out because they appear more two-dimensional at large length scales and more zero-dimensional (point-like) at small length scales (See Figure 4A). If $\Delta s = 0$, the fractal dimension does not change with length scale and the system is self-similar. If Δ*s* is small, then the system is closer to being fractal and the cells are more clustered.

Hotspot Analysis: Another way to see if the cells are clustered or spread out is to determine the fraction of area where there are density 'hotspots', i.e., where the density of cells of a given type, e.g., $CD20^+$ B cells, is above average. To do this analysis, each $CD20^+$ B cell, say, is represented by a two-dimensional Gaussian distribution that represents the local density due to that cell. The width of the Gaussian is $2\sigma$ where $\sigma^2$ is the variance of the Gaussian. We then impose a square lattice of points (with lattice constant *a*) over the image and add the Gaussian weights at each lattice point (see Figure 4B). The resulting sum is the local $CD20^+$ B cell density. We then average the densities over the entire lattice of points and calculate the fraction of lattice points greater than the average for that image. We refer to this as the fraction of the (image) area with hotspots. Note that this fraction is independent of the value of the average B cell density since the hotspots are measured relative to the average density in each particular image. We then average over images and vary $\sigma$. The larger the fraction of hotspots is, the more spread out the cells are.

Our hotspot analysis differs from the Getis-Ord hotspot analysis [58, 60] which requires dividing the image up into regions and depends on the cell counts in neighboring regions. Our hotspot method is completely local in the sense that the cell density at one grid point does not depend on the density at neighboring points.

In applying these techniques to the B cells and killer T cells in the tumors of 37 triple negative breast cancer patients, we found that the fractal dimension difference and the area under the curve of hotspot fraction vs. $\sigma$ is larger for good outcome (see Figure 5), indicating that these cells are more spread out in the cases where the tumor does not recur and more aggregated in cases where there is recurrence.

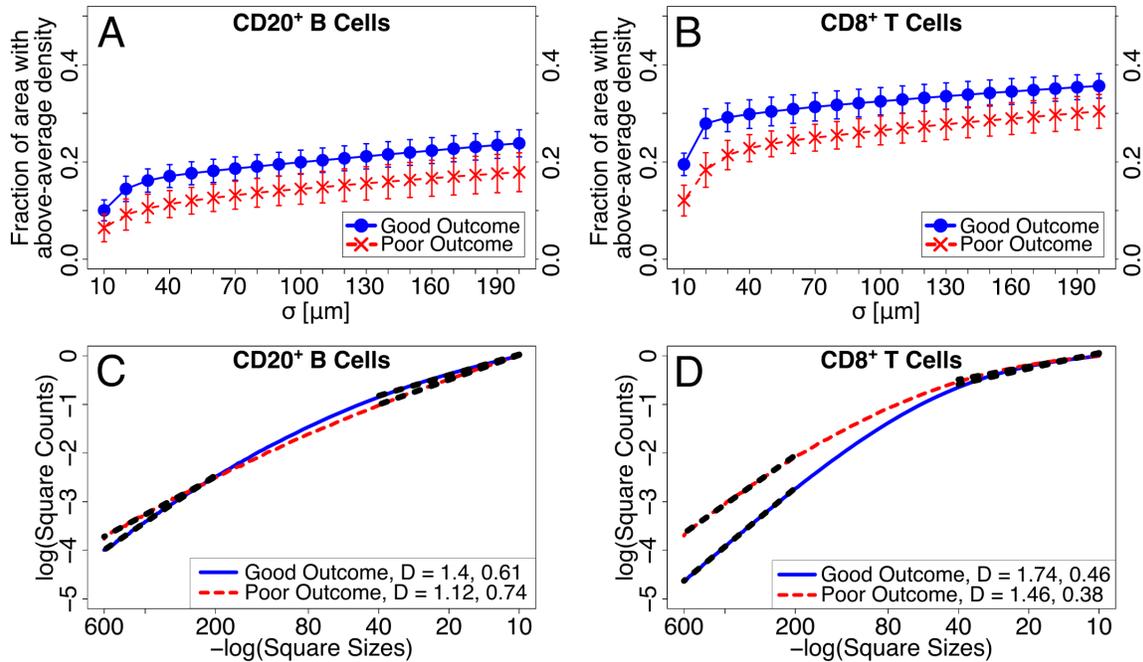

Figure 5 (color online): (A-B) Fraction of area with (A) B cell hotspots (B) $CD8^+$ T cell hotspots vs. σ for good outcome (blue) and poor outcome (red). The error bars correspond to 95% confidence intervals. (C) Log-log plot of the number of squares with at least one $CD20^+$ B cell vs. the inverse box size. (Logarithms are base e.) At long length scales (200-600 microns on the left side of plot), the mean fractal dimension *s* (slope) is 1.4 for good outcome (blue) and 1.12 for poor outcome (red). The p-value for good vs. poor outcome is 0.02 at long length scales. At short length scales (10-40 microns on the right side of the plot), the mean fractal dimension is 0.61 for good outcome and 0.74 for poor outcome. The p-value for good vs. poor outcome is 0.1 at short length scales. The fractal dimension difference Δ*s* = 0.78 for good outcome is greater than Δ*s* = 0.38 for poor outcome. So B cells are more spread out for good outcome. The p-value associated with Δ*s* is $9 \times 10^{-5}$. (D) Log-log plot of the number of squares with at least one $CD8^+$ T cell vs. the inverse box size. At long length scales (200-600 microns on the left side of plot), the mean fractal dimension *s* (slope) is 1.74 for good outcome (blue) and 1.46 for poor outcome (red). The p-value for good vs. poor outcome is 0.0005 at long length scales. At short length scales (10-40 microns on the right side of the plot), the mean fractal dimension is 0.46 for good outcome, 0.38 for poor outcome. The p-value for good vs. poor outcome is 0.2 at short length scales. The fractal dimension difference Δ*s* = 1.28 for good outcome is greater than Δ*s* = 1.08 for poor outcome. So $CD8^+$ T cells are more spread out for good outcome. The p-value associated with Δ*s* is 0.006. (C-D) Black dashed lines show the least squares linear regression fit at long and short length scales. Because different images had different overall areas, we normalized the number n(L) of boxes with 1's by the total number N(L) of boxes with cells in computing the fractal dimension; thus the y-axis values are negative.

*Nearest neighbor (NN) distances between cells of a given type*: An *a priori* obvious way to quantify how spread out the cells of a given type, e.g. B cells, are in an image would be to measure the mean or median NN distances between those cells. However, most B cells are quite close (5-20 μm) to another B cell, so the NN distance just reflects the (inverse of the) local cell density rather than the spatial dispersion at long length scales. However, if the B cells (or cells of a given type) are thinned, then the mean or median nearest neighbor distances can be a good measure of the spatial dispersion of cells [74].

## A Maximum Entropy Approach to Quantifying the Spatial Distribution of Immune Cells

Entropy is a measure of disorder. The greater the number of configurations or the number of ways of arranging things, the greater the entropy is. Mathematically, the entropy is the logarithm of the number of states or ways of arranging things. Since there can be millions of immune cells in an image of the tissue, the number of ways of arranging all these cells and hence, the associated entropy is huge. To reduce the entropy to a manageable value, we took the following approach. We divided the image of the tissue into blocks of size L x L. Each block or matrix was subdivided further into 3 x 3 squares like tic-tac-toe. As before, for each square, we then ask a binary (yes-no) question, e.g., "Is there at least one T cell in the square?" If the answer is yes, we assign a 1 to that square. If the answer is no, we assign a 0 to the square. There are a total of $2^9$ or 512 possible states. So the entropy for total randomness (all states equally likely) is $S_0 = \ln(512)$. We chose 3 x 3 matrices in order to obtain meaningful statistics, i.e., we want the total number of possible states to be much less than the number of samples (statistical trials). Let $P(x_i)$ be the probability of finding the *i*th configuration where *i* varies from 1 to 512. Then the entropy is

$$S = -\sum_{i=1}^{512} P(x_i) \ln\left[P(x_i)\right] \tag{1}$$

Maximum Entropy: Maximum entropy is a widely used technique to find the probability $P(x_i)$ subject to the constraint that it produces the observed expectation values of various quantities of the distribution [77, 78]. It is a minimalist approach in that it does not impose any other constraints or assumptions. It works in the following way. Let $f_\mu(x_i)$ be the μth measureable quantity when the system is in the state $x_i$. The observed expectation value is $<f_\mu>_{expt}$. The constraint is that the probability $P(x_i)$ satisfy $\sum_i P(x_i) f_\mu(x_i) = \langle f_\mu \rangle_{expt}$. To do this we maximize the entropy functional:

$$\tilde{S}\left[P(x)\right] = -\sum_i P(x_i) \ln P(x_i) - \sum_{\mu=0}^{K} \lambda_\mu \left[\sum_i P(x_i) f_\mu(x_i) - \langle f_\mu \rangle_{expt}\right] \tag{2}$$

where $\lambda_\mu$ are the Lagrange multipliers that are adjusted so that the constraints are satisfied. Maximizing $\tilde{S}$ yields the probability $P(x_i) = \frac{1}{Z}\exp\left[-\sum_{\mu=1}^{K}\lambda_\mu f_\mu(x_i)\right]$ where Z is the normalization constant. The entropy is then given by $S = -\sum_i P(x_i)\ln[P(x_i)]$.

Moments of the Distribution, Probabilities and Entropies [79]: The observables that set the constraints are the moments of the distribution. The zeroth moment is the normalization condition on the probability. In our case, the first moment is the average matrix occupancy <p>, i.e., the average fraction of squares that have a '1' in a matrix of 3 x 3 squares. If $P_1(x_i)$ is normalized and is subject to the constraint that it gives the right value of <p>, then this gives the first order entropy $S_1$. The connected information of order 1 is defined by $I_1 = S_0 - S_1$ and tells us how much the occupancy or density reduces the entropy from the completely random case [79]. (Note that this definition of occupancy is slightly different from our earlier description where we placed a grid of squares covering the image of the tissue, and defined the occupancy to be the fraction of squares with 1's.)

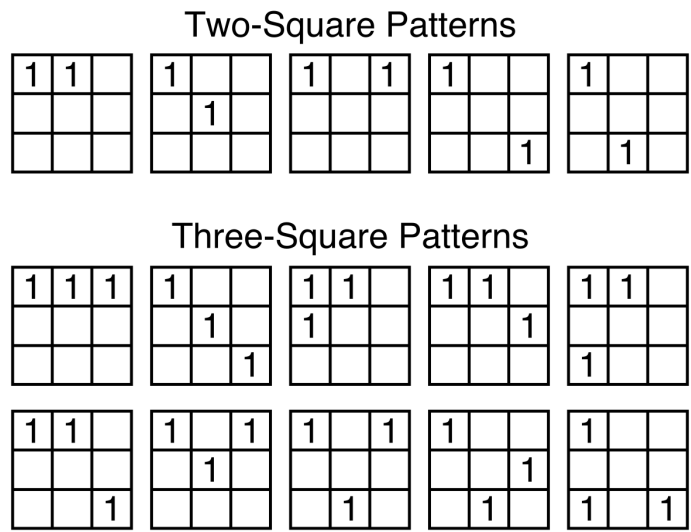

Figure 6: The five two-square and ten three-square patterns possible for a 3 x 3 matrix, ignoring translations, rotations, and reflections. The two-square (three-cell) pattern has two (three) 1's and the blank squares can be either 1 or 0. Each pattern corresponds to a state. Rotations, translations and reflections of a given pattern are counted as part of the same state. The fraction of times that a given state µ occurs equals <$f_\mu$>$_{expt}$ in Eq. (2).

The second moment consists of the average occurrence rates of various patterns with two squares that have 1's. Ignoring translations, rotations and reflections, there are 5 types of such patterns in a matrix of 3 x 3 squares as shown in Figure 6. If $P_2(x_i)$ is normalized and gives the right value of <p> as well as those of the 2-square occurrence rates, then this gives the second order entropy $S_2$. The connected information of order 2, $I_2 = S_1 - S_2$, tells us how much the constraints of the second moment reduces the first order entropy. Similarly, the third moment corresponds to 10 different arrangements of 3 squares that have '1' in a 9-square matrix, ignoring translations, rotations and reflections as shown in Figure 6. $P_3(x_i)$ gives the third order entropy $S_3$, and the connected information of order 3 is given by $I_3 = S_2 - S_3$. The multi-information is defined by $I_N = S_0 - S_N$ where $S_N$ is given by using the actual histogram distribution found from the data.

Relation of Occupancy and $I_1$: We note that $I_1$ is related to occupancy in the following way for a 3 x 3 matrix of squares. Let the occupancy p be the probability that a square will have a '1'. Let q = (1-p) be the probability that a square will have a '0'. Let $x_i$ denote a given state or arrangement of 1's and 0's in a 3x3 matrix. Then the probability $\tilde{P}_{n,p}(x_i)$ of a given state with n 1's is given by $\tilde{P}_{n,p}(x_i) = p^n q^{9-n}$. The first order entropy just depends on the occupancy and is given by

$$S_1 = -\sum_{i=1}^{512} \tilde{P}_{n,p}(x_i) \ln\left[\tilde{P}_{n,p}(x_i)\right] = -\sum_{n=0}^{9} \left[\frac{9!}{n!(9-n)!}\right] \tilde{P}_{n,p} \ln\left[\tilde{P}_{n,p}\right] \quad (3)$$

where we used the binomial coefficient to reduce the number of terms in the sum from 512 to 10 since the binomial coefficient gives the number of ways to arrange n 1's in a 3 x 3 matrix. The connected information $I_1 = S_0 - S_1$ where $S_0 = \ln(512) = 6.238$. Note that $I_1 = 0$ at p = q = 0.5 because $\tilde{P}_{n,p=0.5} = 1/512$. We can explicitly evaluate Eq. (3) and compute $I_1$ versus occupancy. This is plotted in Figure 7. Notice that for occupancy p > 0.5, the greater the occupancy, the higher $I_1$ is, whereas for p < 0.5, the greater the occupancy, the lower $I_1$ is.

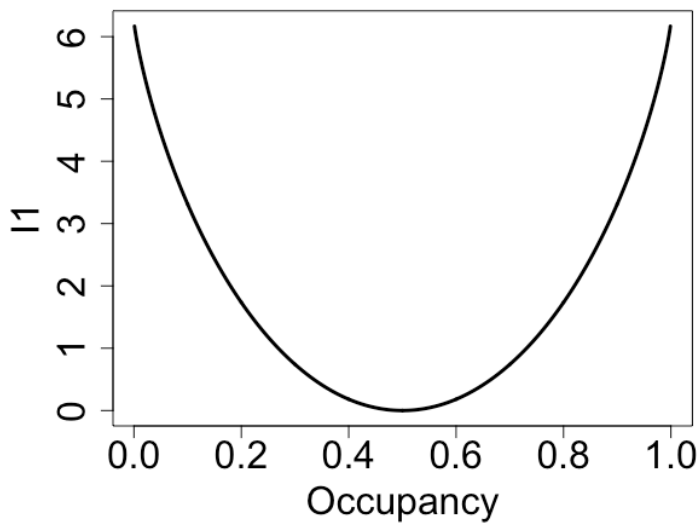

**Figure 7: $I_1$ vs. occupancy. $I_1 = 0$ and has a minimum when the occupancy = 0.5 because that is where there is a 50% chance of having a 0 or a 1 in each square. So the entropy $S_0 = S_1$.**

*Kullback-Leibler Divergence:* Another way to compare the different probability distributions, e.g., $P_1$ and $P_2$, is through the Kullback-Leibler (KL) divergence [80, 81] (or relative entropy):

$$D_{12} = -\sum_i P_1(x_i) \ln\left(\frac{P_2(x_i)}{P_1(x_i)}\right)$$

This is a measure of how different a probability distribution $P_2$ is from a reference (*a priori*) probability distribution $P_1$. If the distributions are identical, then $D_{12} = 0$; otherwise $D_{12}$ is always positive.

## Application of Maximum Entropy to Breast Cancer

### Triple Negative Breast Cancer Patient Cohort

We have used our maximum entropy approach to analyze the spatial distribution of various types of TILs in 2D images of tumor-infiltrating lymphocytes (TILs) in immunohistochemistry-based images of primary tumor tissue from 37 patients with triple negative breast cancer (TNBC) prior to any treatment. In this study, we defined patients with no recurrence within 5 years after surgery as good clinical outcome (n=24) and patients who had recurrence within 3 years after surgery as poor clinical outcome (n=13). All the patients were subsequently treated with standard chemotherapy; some also had radiotherapy.

### Preparation of Images: Multispectral Staining of Different Cell Types

To analyze the spatial distribution of various types of TILs, we first identified the locations of cells in 2D images of archived formalin fixed paraffin embedded (FFPE) tumor tissues from 37 TNBC patients. Each cell type was immunostained and labeled with a different color chromophore. This means that the tissue is treated with an antibody that binds to a protein that is specific to a certain cell type. For example, the protein CD3 is expressed on all types of T cells but not on B cells. So if we attach a green colored molecule or chromophore to CD3 antibodies, then T cells that bind the antibodies will appear green. Each image was restricted to tumor-associated regions and excluded necrotic and fibrotic areas determined by a pathologist. (Necrotic tissue consists of dead cells. Fibrotic tissue has excess fibrous connective tissue, e.g., scar tissue.)

## RESULTS
### Entropies vs. Length Scale

In Figure 8, we plot $S_1$, $S_2$, $S_3$ and $S_N$ versus matrix size for good and poor outcome for B cells and CD8$^+$ T cells. (Recall that a matrix has 3x3 squares.) Notice that there is a distinct difference between good and poor outcome, though the shape of the curves is similar because the densities of B cells is comparable between good ($3.0 \times 10^2$/mm$^2$) and poor ($2.3 \times 10^2$/mm$^2$) outcome. If we integrate under the curves to get the area under the curve (AUC), then good outcome has a larger AUC for B cells compared to poor outcome. For example, the B-cell AUC for $S_N$ is 31 for good outcome compared to 22 for poor outcome. This difference is significant (p-value=0.01) and reflects the fact that the B cells are more spread out for good outcome and more aggregated for poor outcome. However, there is no significant difference between the good and poor outcome AUC for CD8$^+$ T cells because the curves cross. The difference between good and poor outcome in the shape of the curves for CD8$^+$ T cells is due the significant difference in densities between good ($4.5 \times 10^4$/mm$^2$) and poor ($2.1 \times 10^4$/mm$^2$) outcome. The maximum in $S_1$ vs. matrix size occurs at the matrix size where the probability p of having at least one CD8$^+$ T cell in a square is closest to 0.5. Recall that the case of $p = q = 0.5$ gives the maximum entropy $S_0$. Since good outcome has a higher CD8$^+$ T cell density, the entropy curves reach a maximum at a smaller matrix size compared to poor outcome.

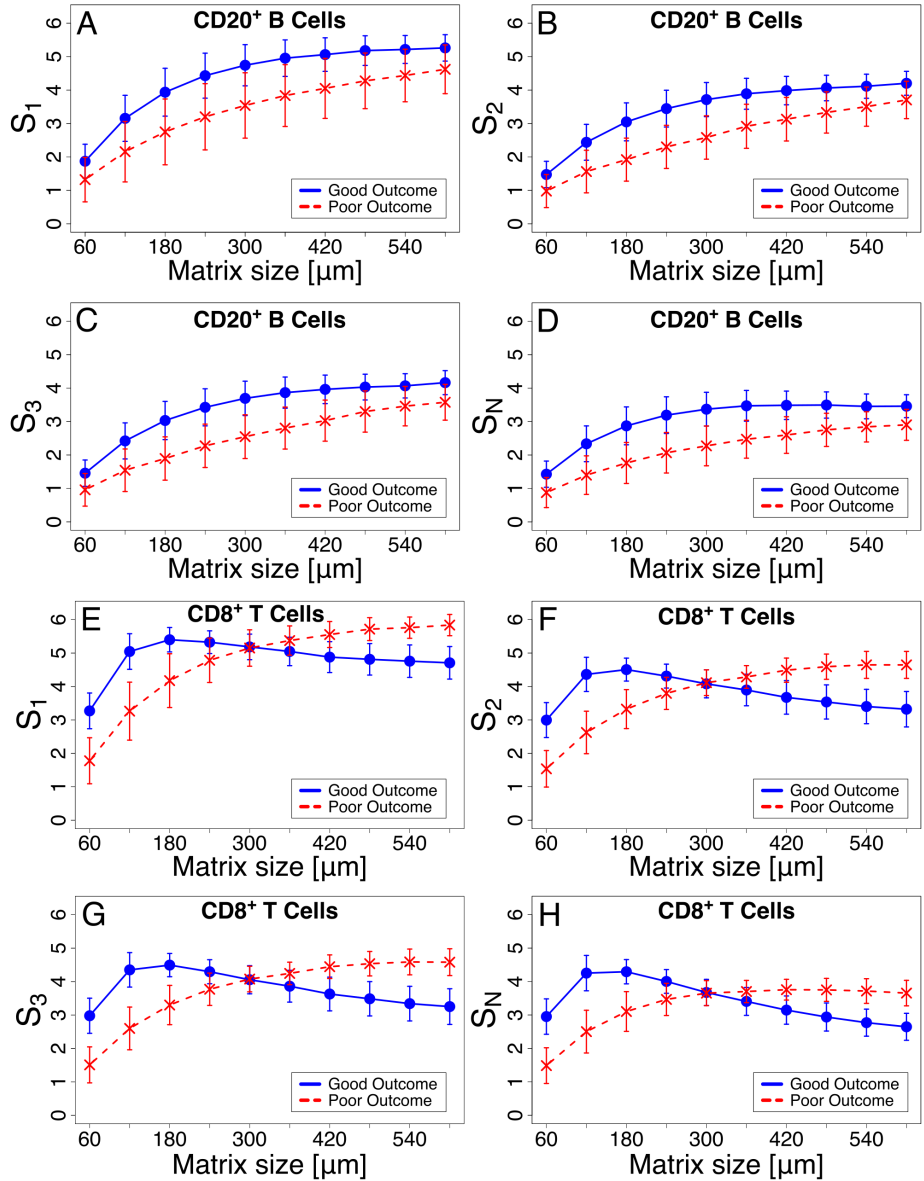

**Figure 8 (color online):** Entropies $S_1$, $S_2$, $S_3$ and $S_N$ vs. matrix size for (A-D) CD20$^+$ B cells and (E-H) CD8$^+$T cells. Blue solid lines are for good outcome and red dashed lines are for poor outcome. The error bars represent 95% confidence intervals. (A) $S_1$ (B) $S_2$ (C) $S_3$ (D) $S_N$ (E) $S_1$ (F) $S_2$ (G) $S_3$ (H) $S_N$

Notice that for a given cell type and outcome, the entropy curves in Fig. 8 are similar to those for $S_1$. This is because $S_1$ dominates the entropy. This is clearly seen in Figs. 9 and 10 where the connected information moments ($I_1$, $I_2$, $I_3$, and $I_N$) and KL divergences ($D_{01}$, $D_{12}$, $D_{23}$, and $D_{0N}$) are plotted for B cells and CD8 T cells. One can see that similar information is conveyed for the pairs ($I_1$ and $D_{01}$), ($I_2$ and $D_{12}$), etc. Both connected information moments and KL divergences are ways to compare two distributions. $I_1$ and $D_{01}$ differ between good and poor outcome but there is no noticeable dependence on outcome for $I_2$, $I_3$, $D_{12}$, and $D_{23}$. In fact, $I_3$ and $D_{23}$ are negligible. $I_2$ and $D_{12}$ are nonzero and reflect clustering of the cells since the cells need to be in the same matrix. $I_N$ is similar to $I_1$ because it is dominated by $I_1$, and $D_{0N}$ is similar to and dominated by $D_{01}$. The fact that the difference between good and poor outcomes is only noticeable for the first moment ($I_1$ and $D_{01}$) is why we focused on quantities described earlier like occupancy, fractal dimensions, and fraction of area with density hotspots.

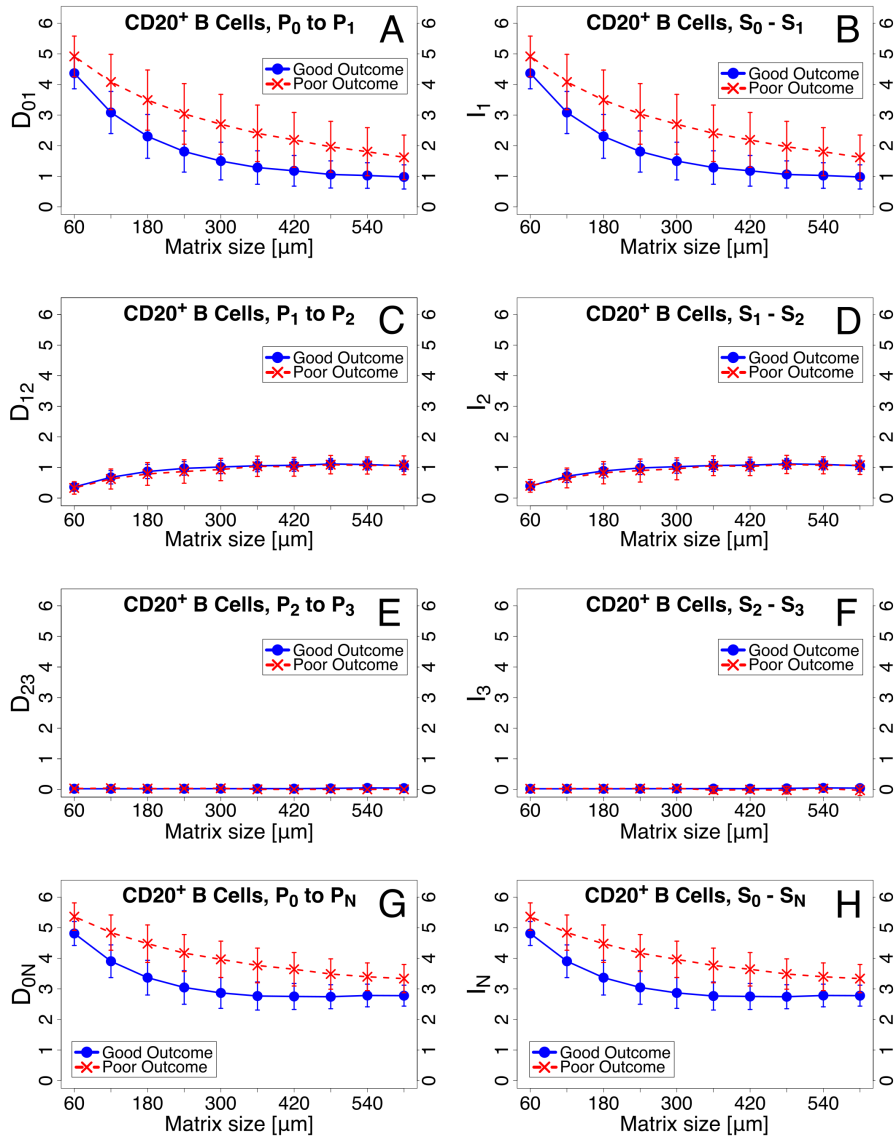

Figure 9 (color online): Information moments and Kullback-Leibler divergences vs. matrix size for CD20[+] B cells. Blue solid lines are for good outcome and red dashed lines are for poor outcome. The error bars represent 95% confidence intervals. (A) $D_{01}$ (B) $I_1$ (C) $D_{12}$ (D) $I_2$ (E) $D_{23}$ (F) $I_3$ (G) $D_{0N}$ (H) $I_N$.

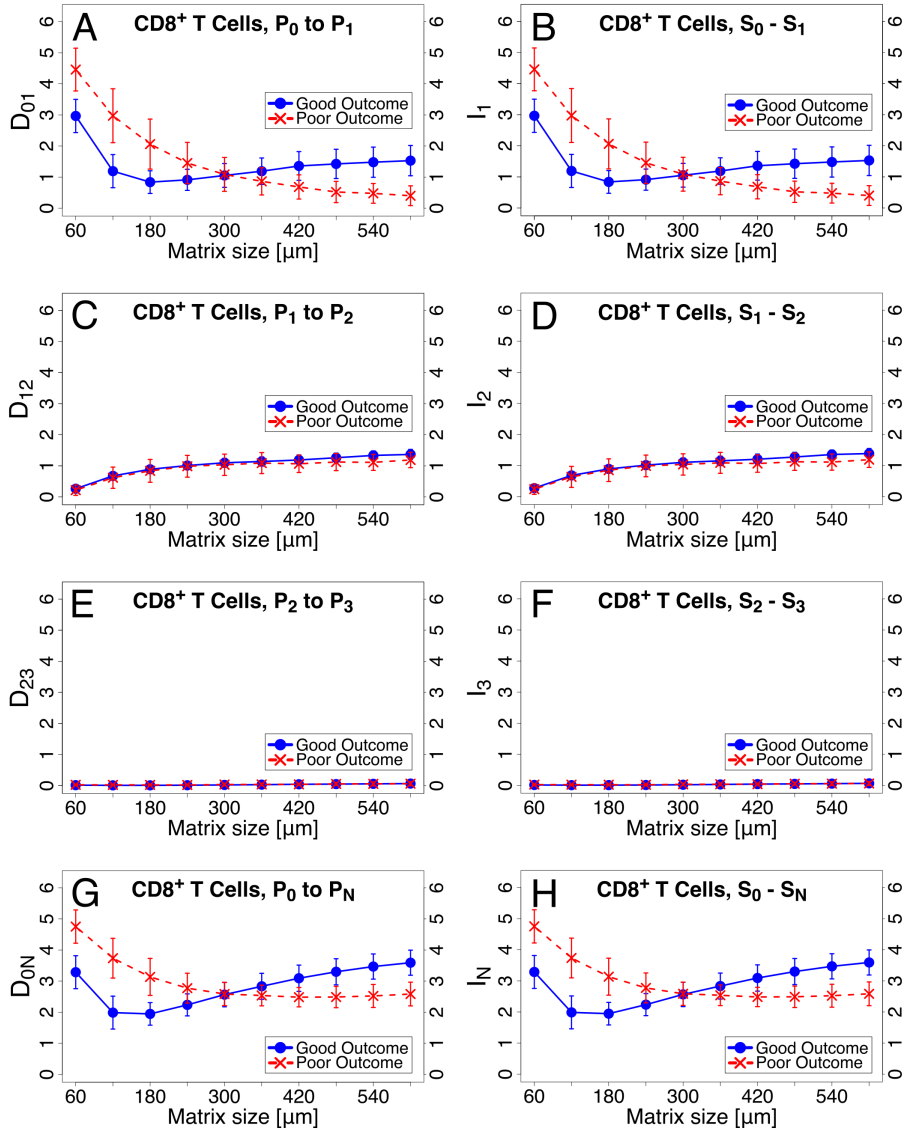

Figure 10 (color online): Information moments and Kullback-Leibler divergences vs. matrix size for CD8+ T cells. Blue solid lines are for good outcome and red dashed lines are for poor outcome. The error bars represent 95% confidence intervals. (A) $D_{01}$ (B) $I_1$ (C) $D_{12}$ (D) $I_2$ (E) $D_{23}$ (F) $I_3$ (G) $D_{0N}$ (H) $I_N$.

## DISCUSSION

We have seen that going beyond the density of immune cells, e.g., B cells or T cells, and looking at their spatial distribution can reveal differences associated with clinical outcome, i.e., whether or not cancer recurs. Unlike previous approaches to quantify the spatial distribution of cancer and immune cells, we have developed techniques that span a range of length scales. Occupancy, fractal dimension difference and the fraction of area with density hotspots are useful for determining to what degree cells are spread out or aggregated. In our maximum entropy approach, $S_1$, $I_1$ and $D_{01}$ are sensitive to the density of cells while higher moments, e.g., $S_2$, $I_2$ and $D_{12}$, are a way to ascertain the spatial correlation of cells on different length scales.

Our findings raise a number of important questions about the spatial distribution of TILs. First, what determines the spatial distribution of TILs? Certainly, chemical signals, i.e., cytokines and chemokines help to dictate where immune cells go. In addition, we speculate that the somewhat fractal spatial distribution of TILs may arise from the branching trajectories of the B and T cells as they patrol the tissue. Branching structures such as trees and plant roots are self-similar, and hence fractal, because they look the same over a range of length scales, i.e., over a range of magnifications. It may be that the paths that B and T cells travel along have a branching structure because these cells have to go around physical obstacles such as other cells, blood vessels, and collagen fibers. In addition, T cells are known to follow along the outside of blood vessels and collagen fibers [82, 83] which can have a branching architecture.

Second, why do the spatial distributions of B and T cells differ between good and poor clinical outcome? In particular, one could also ask why B and T cells are more spread out for good clinical outcome as found in Ref. [74]. One possibility is that the difference in spatial distribution reflects differences in the spatial topography (obstacles) of the tumor microenvironment. In FFPE tissue sections, we performed immunohistochemical staining of collagen I which did not yield differences in the spatial distribution of collagen I between good and poor outcome (data not shown). This indicates that differences in physical topography may not be the explanation.

It may be that when they are more spread out, these types of cells are better at immune surveillance in good outcome patients. For example, antigen-presenting B cells may be more likely to acquire and successfully present cancer neo-antigens to helper T cells if they are able to interact both with cancer cells and other immune cells by spreading out. In addition, cytokine secreting B cells may cover a larger area with signaling molecules. In either case, an even spatial distribution is a more effective strategy for B and T cell surveillance than sequestration in small parts of the tumor. It may also be that B and T cells in good outcome patients are more responsive to chemokines secreted by tumor cells and are more successful in finding their cognate (matching) antigens within the tumor microenvironment as in a recently published study [84].

Third, it is rather odd *prima facie* that we are able to give a prognosis with any degree of accuracy based on tumor tissue that has been removed from a patient. The cells in the surgically removed tissue are no longer in the patient, yet their spatial distribution can be used to predict whether or not the cancer will recur with an accuracy of 60-80%. The reason for this is not understood. It may be that spatially dispersed B and T cells are associated with more immune engagement and the production of more memory B and T cells, leading to a better long-term outcome.

The statistical techniques that we have developed to quantify spatial distributions are novel. Unlike previous efforts to analyze spatial distributions at a single length scale, we have examined how the spatial distribution varies with length scale and how that can shed light on whether the cells are clustered or spread out spatially. On a more general level, these approaches are flexible and can be applied to a broad spectrum of problems. There are straightforward extensions of these approaches, e.g., to quantify the spatial distributions of various kinds of cells in metastatic tumors or to ascertain which patients are good candidates for various therapeutic treatments. However, one can go beyond simply the spatial distribution of cells or discrete entities. In determining the occupancy, fractal dimension and the difference in fractal dimensions, we laid down a grid of squares and asked a binary yes-no question of each square. In this paper, we asked questions like "Is there at least one $CD20^+$ B cell in this square?" However, in other contexts, one may be interested in other questions such as that of co-localization. For example, one could ask "Does the square have at least one T cell that is within 50 microns of a dendritic cell?" or "Does the square have at least one B cell and one T cell within 25 microns of a blood vessel?" Furthermore, these approaches could be applied to image analysis outside of biology, e.g., the distribution of galaxies in astrophysics.

In conclusion, we have presented novel techniques to quantify the spatial distribution of point-like objects. In applying these techniques to B and killer T cells in tumors, we found that the spatial arrangement of these immune cells is strongly correlated with clinical outcome, i.e., breast cancer recurrence. This highlights the importance of these immune cells in cancer and raises new questions about their role in preventing cancer.

**Acknowledgements**
We thank Dr. Ching Ouyang for helpful discussions. CCY thanks Dr. Arnold Levine and Dr. Larry Norton for helpful discussions. This work was supported by Stand Up To Cancer, The V Foundation, and the Breast Cancer Research Foundation. The work of CCY and JCW was also supported in part by the Cure Breast Cancer Foundation. The work of CCY was performed in part at the Aspen Center for Physics, which is supported by National Science Foundation grant PHY-1607611.